\documentclass[conference]{IEEEtran}
\IEEEoverridecommandlockouts
\usepackage{cite}
\usepackage{amsmath,amssymb,amsfonts}
\usepackage{algorithmic}
\usepackage{graphicx}
\usepackage{textcomp}
\usepackage{xcolor}
\usepackage{listings}
\usepackage{enumerate}
\usepackage{todonotes}  
\usepackage{placeins}

\usepackage{url}
\usepackage{breakurl}
\usepackage[breaklinks]{hyperref}

\usepackage{caption} 
\usepackage{stfloats}

\newcommand{\smallttt}[1]{\small\texttt{#1}\normalsize}

\def\BibTeX{{\rm B\kern-.05em{\sc i\kern-.025em b}\kern-.08em
    T\kern-.1667em\lower.7ex\hbox{E}\kern-.125emX}}
\begin{document} 

\title{A Flexible Cell Classification for ML Projects in Jupyter Notebooks
}
% Newpstate of the art solutions ...?

\author{\IEEEauthorblockN{Miguel Perez}
\IEEEauthorblockA{\textit{Research Group Software Construction} \\
\textit{RWTH Aachen University}\\
Aachen, Germany \\
miguel.perez@rwth-aachen.de}
\and
\IEEEauthorblockN{Selin Aydin}
\IEEEauthorblockA{\textit{Research Group Software Construction} \\
\textit{RWTH Aachen University}\\
Aachen, Germany \\
aydin@swc.rwth-aachen.de}
\and
\IEEEauthorblockN{Horst Lichter}
\IEEEauthorblockA{\textit{Research Group Software Construction} \\
\textit{RWTH Aachen University}\\
Aachen, Germany \\
lichter@swc.rwth-aachen.de}
}

\maketitle

\begin{abstract}
Jupyter Notebook is an interactive development environment commonly used for rapid experimentation of machine learning (ML) solutions. Describing the ML activities performed along code cells improves the readability and understanding of Notebooks. Manual annotation of code cells is time-consuming and error-prone. Therefore, tools have been developed that classify the cells of a notebook concerning the ML activity performed in them.  However, the current tools are not flexible, as they work based on look-up tables that have been created, which map function calls of commonly used ML libraries to ML activities. These tables must be manually adjusted to account for new or changed libraries.

This paper presents a more flexible approach to cell classification based on a hybrid classification approach that combines a rule-based and a decision tree classifier. We discuss the design rationales and describe the developed classifiers in detail. We implemented the new flexible cell classification approach in a tool called \textsc{JupyLabel}. Its evaluation and the obtained metric scores regarding precision, recall, and F1-score are discussed. Additionally, we compared \textsc{JupyLabel} with \textsc{HeaderGen}, an existing cell classification tool. We were able to show that the presented flexible cell classification approach outperforms this tool significantly.

\end{abstract}


\section{Introduction}

Jupyter Notebook is the tool most commonly used by data scientists when experimenting with Machine Learning (ML) solutions \cite{10.1145/3392826                     }. It is an interactive programming environment where code is written and executed in a document-like structure. The produced Jupyter Notebook document is usually called a \textit{notebook}. The output of code cells, such as text output or visualizations, is displayed below the executed cell. By adding descriptions to the code, data scientists can describe insights about the data and how solutions work in narrative form. This makes notebooks very convenient for exploratory programming.  

Nevertheless, the unstructured and iterative approach of exploratory programming often leads to disorganized notebooks that do not adhere to best practices. For instance, in a large-scale case study, Pimentel et al. \cite{pim} analyzed around 1.2 million notebooks on \textsc{GitHub} and found that only 24.11\% of the notebook results were reproducible.

To address this problem, Rule et al. \cite{rule2018simple} established several rules for better reproducibility and collaboration in notebooks. The authors emphasize the importance of documentation and story-telling within a notebook, especially describing what a cell does.

However, of the 1.2 million notebooks analyzed in \cite{pim}, 30.93\,\% do not contain a single Markdown cell, and 50\,\% of the notebooks that contain Markdown cells do not contain any meaningful information about those cells, making the Markdown cells useless for documentation in these cases. Interviews conducted by Rule et al. \cite{10.1145/3173574.3173606                        } and Wenskovitch et al.\cite{8973385} suggest that Markdown cell documentation is not added because time is short or other tasks have higher priority.

Tools have been developed to facilitate the understanding of notebooks that classify cells according to ML activity performed and insert corresponding heading labels into the notebook cells.

This paper is organized as follows. In Section \ref{sec:related-work}, we overview existing approaches to code classification in general and cell classification in notebooks. Then, in Section \ref{sec:statement}, we formulate the problem to be solved and the goals we aim to achieve. An explanation of all ML activities according to which the cells of a notebook are to be classified follows in Section \ref{sec:labels}. Next, Section \ref{sec:design} presents the design of the proposed flexible cell classification approach and a description of the developed classifiers. Then, in Section \ref{sec:JupyLabel}, we briefly describe the architecture and components of the developed tool called \textsc{JupyLabel}. In Section \ref{sec:eval}, we present the evaluation performed and discuss the results. In addition, \textsc{JupyLabel} is compared with the existing cell classification tool \textsc{HeaderGen}. The paper concludes with a summary of the contributions and an outlook on future research directions in Section \ref{sec:conclusion}.

\section{Related Work}
\label{sec:related-work}
Because the classification of code cells in notebooks can be traced back to source code classification, we first present such well-known approaches. Then, we discuss already published tools for classifying notebook cells for ML projects.

\subsection{Approaches to Source Code Classification}
Regarding source code classification, extensive research has already been done to detect different patterns, including code smells \cite{codesmelldet} or application domains \cite{ugurel2002s}. Various techniques have addressed this problem, like binary, multi-class, and multi-label text classification methods. 

In most of these approaches, identifying each specific pattern is considered a binary classification problem, as shown by works such as \cite{azeem19, khomh09, Maneerat11}. In doing so, each pattern is associated with its binary classifier, resulting in an ensemble of classifiers, each specialized in detecting a particular one.

To carry out a binary classification, in the simplest case, one can first define rules based on domain knowledge to decide to which class an object to be classified belongs. In more complex cases, these rules need to be learned from analyzing data. In the following, the rule-based, ML-based, and hybrid classifiers are discussed in more detail.

\subsubsection{Rule-based Classifier}
When dealing with less complex classification tasks, rule-based classifiers are simple and effective \cite{li2014rule}. For instance, detecting the \textit{long method} code smell can be effectively accomplished using straightforward rules, like predefined thresholds for lines of code or parameters per method. 

These rules include if-then statements, manually constructed decision trees, or look-up tables. The simplicity leads to high interpretability of classification results.  

However, it's important to note that this type of classifier is static. When the environment changes, altering the underlying assumptions used to define rules, the rules must be manually updated accordingly.

\subsubsection{ML-based Classifier}
In more complex cases, the definition of explicit rules is more complicated. That is why various supervised ML classifiers can be applied to binary text classification problems.

The choice of algorithm depends on the classification problem on hand, e.g., Naive Bayes works with the assumption that the probability of occurrence of each word is independent of the other words. The opposite is assumed for algorithms like support vector machines, neural networks, and decision tree (DT) algorithms. 

A case study by Amorim et al. \cite{amorim2015experience} found that DT algorithms outperform other approaches regarding accuracy in detecting code smells in most scenarios. In addition, DT algorithms are significantly less computationally expensive. However, it is essential to acknowledge that these algorithms excel in classification speed but are also more susceptible to overfitting issues. To mitigate this problem, ensemble methods combining multiple DTs, such as bagging, boosting, or random forests, are used \cite{DTcomp}. 

In contrast to rule-based classifiers, ML-based classifiers can be retrained on newer data and thus adapt to changes in the environment more easily.

\subsubsection{Hybrid Classifier}
A comprehensive comparison of ML techniques for code smell detection emphasizes the advantages of integrating rule-based classifiers alongside ML classifiers \cite{arcelli2016comparing}. ML classifiers, such as DT classifiers, alleviate the cognitive burden on developers by automating the process of rule discovery and generation through extensive data analysis. If the rule is simple and easy to define, it is thus not necessary to utilize an ML classifier.

\subsection{Cell Classification Approaches \& Tools}
With \textsc{HeaderGen}, Venkatesh et al. \cite{venkatesh2023enhancing} present an approach and a tool that adds categorical Markdown headers to code cells based on a taxonomy of ML activities. In this approach, the authors create a library-to-ML-activity look-up table in which function calls of commonly used ML libraries, such as \textit{scikit-learn} or \textit{pandas}, are mapped to ML activities. Based on this, an analysis of function calls per cell provides the corresponding labels. Since data operations can be performed without function calls, they extend their approach by matching patterns consisting of assignment or data access operands. 

A user study showed the usefulness of \textsc{HeaderGen} for enhanced navigation and comprehension. 

The accuracy of \textsc{HeaderGen} was evaluated using 15 notebooks taken from Kaggle \cite{Quaranta_2021}. The generation of headers achieved a precision of 82.2\,\% and a recall rate of 96.8\,\% compared to manually created headers by experts. Regarding the precision scores, it should be noted that the sample of 15 notebooks was selected according to specific criteria. These include that a selected notebook must contain popular ML libraries. This circumvents a known limitation of \textsc{HeaderGen}, namely classifying cells that use lesser-known or custom ML libraries. In addition, the chosen notebooks are already very well structured, having only a few lines of relatively simple code per cell. 

A similar approach was developed by Jiang et al. \cite{jiang}. The authors create an API-to-ML-activity look-up table by mapping API calls from popular ML libraries to activity labels. In their evaluation, they used 50 notebooks with a total of 1208 cells and achieved 75\,\% accuracy.

\section{Problem Statement and Research Goals}
\label{sec:statement}
Because existing approaches and tools for classifying notebook cells are based on the identifiers of commonly used ML libraries, they depend directly on them.  

Accordingly, adding new functions to existing libraries, renaming or deprecating them requires constantly updating the look-up tables. The same applies if user-defined or less well-known ML libraries are used.

These approaches can only be applied reliably with an up-to-date look-up table. However, this requires continuous maintenance efforts because new releases of all supported ML libraries must be monitored to update the look-up table. The developers of \textsc{HeaderGen} \cite{venkatesh2023enhancing} also mention that their look-up table is still in progress to cover more ML libraries. 

Hence, these approaches lack flexibility regarding the used environment, i.e., the used ML libraries and their versions, as they change frequently.

Based on the points above, the research presented in this paper aims to address the following  goals:
\begin{itemize}
    \item[G1:] Design a flexible cell classification approach for ML projects in Jupyter Notebook that overcomes the limitations of existing ones without losing accuracy.
    \item[G2:] Provide a helpful and fast tool for applying the new cell classification approach on real-world notebooks.
\end{itemize}
Before presenting the design of a flexible cell classification approach, we briefly introduce the set of \textit{conceptual ML activities} used for classification.

\section{ML Activity Classes Considered}
\label{sec:labels}
For a notebook cell to be classified, the conceptual ML activity (activity for short), usually implemented by several concrete ML tasks, that is performed in the cell must be determined. We reduce this task to a multi-label classification problem with a set of activity labels. Therefore, in this section, we provide a set of suitable activity labels, which are essentially based on the phases and tasks of the CRISP-DM model \cite{wirth2000crisp}, defining the phases: (1) Business Understanding, (2) Data Understanding, (3) Data Preparation, (4) Modeling, (5) Evaluation, and (6) Deployment.

To understand (2) and prepare (3) the data, the following activities are performed:
\begin{itemize}
    \item \textit{ingest data:} The initial data is read and loaded. It serves as the source for the data used in the ML model.  
    \item \textit{validate data:} The ingested data gets validated. On one hand, this can be done programmatically, e.g., the data format is checked using \textit{assert} statements. On the other hand, manually, i.e., by including code statements that don't alter data or models but are used to visually inspect and informally validate data by printing information.         
    \item \textit{process data:} The validated dataset is used and processed, e.g., sorted or split into test and validation data. After that, the processed dataset can be further used, e.g., to visualize the data or build the ML model.      
\end{itemize}
To construct and evaluate the ML model (4 \& 5) and to deploy it (6), these activities are usually needed: 
\begin{itemize}
    \item \textit{train model:} All ML model training-related tasks, e.g., model instantiation, initialization, or configuration of parameters, are mapped to this activity.
    \item \textit{evaluate model:} The ML model is evaluated using metrics like precision, recall, or F1-score, using separate datasets for evaluation purposes. Analysis of the metrics may show that further optimization is required, like hyperparameter tuning or a completely different model architecture. Additionally, any prediction or inference tasks are mapped to this activity.
    \item \textit{transfer results:} This activity includes all kinds of deployment tasks, like deploying the ML model in a repository, production, exporting the data to a data format, or exporting pre-trained models.  
\end{itemize} 
In addition, other activities must usually be performed. These are:
\begin{itemize}
    \item \textit{setup notebook:} To set up a notebook's environment, libraries are imported, magic commands are executed, and constants are declared, e.g., for hard-coded paths. 
    \item \textit{visualize data:} It includes tasks (e.g., data plotting) to create visualizations like diagrams or heatmaps. It is often necessary to pre-process the data to create plots and charts.
\end{itemize}

\section{Design of the Cell Classification Approach} 
\label{sec:design}
Since a flexible cell classification approach must not depend on ML library information, it must only consider the information contained in the cells themselves.

A manual analysis of notebooks revealed that some activities can be classified based on the cell information and the code elements in the cells. For these cases, we designed a more or less simple rule-based classifier. 

However, to classify all types of activities with high precision, the rule-based classifier must be complemented by an ML-based classifier, in our case, a DT-based classifier.

Therefore, we designed a hybrid cell classification approach to benefit from a rule-based classifier's simplicity and the DT-based classifier's power.

\subsection{The Rule-based Classifier}
\label{sec:design-RBC}
In the JSON representation of a notebook, every cell contains meta information, e.g., \textit{cell\_type}, or \textit{output\_type}. Based on such meta information and code elements, heuristics can be formulated for the following activities:
\begin{itemize}
    \item A cell often realizes a \textit{setup notebook} activity,
    \begin{itemize}
        \item if it contains \textit{import} statements to configure and set the environment,
        \item if it contains \textit{constant} declarations,
        \item if it contains \textit{magic commands} used to interact with the IPython kernel directly, e.g., to install additional packages.
    \end{itemize}
    \item A cell often realizes a \textit{visualize data} activity, 
        \begin{itemize}
        \item if the \textit{output\_type} of a cell is \textit{display\_data}, in this case, the output of a cell is a plot.
    \end{itemize}    
    \item A cell often realizes a \textit{validate data} activity,
    \begin{itemize}
         \item if the cell or its output contain keywords like \textit{assert}, \textit{verify}, or \textit{check},
        \item if it contains a single connected string in a line, which is not a method call but requests the value of an object, such as in \smallttt{df} or \smallttt{df.columns},
        \item if it contains \textit{print} statements.
    \end{itemize}
\end{itemize}

These heuristics can then be transformed into respective rules, which are applied by the rule-based classifier.

\subsection{The DT-based Classifier}
Since DT algorithms are often used for code classification, we broke the problem into eight binary classification problems, one for each activity. Therefore, we developed eight DT models trained to detect exactly one activity. 

\subsubsection{Feature Extraction} 
When using ML for classification tasks, a crucial step is \textit{tokenizing}. The \textit{CountVectorizer} implements tokenizing and occurrence counting and creates a dictionary from a given corpus \cite{scikitCountVec}. After that, each sentence is transformed into a vector containing the number of occurrences of each word from the dictionary. 

To ensure that code operators frequently used in \textit{process data} activities such as \smallttt{[} and \smallttt{]} as well as the assignment operator \smallttt{=} are also captured by the CountVectorizer, we have modified the regular expression used as the default token pattern accordingly. Therefore, the tokenizer applies the regular expression $[a-zA-Z]\{1,\}|[=[\backslash]]$ (matching one or more letters or an equal sign or a square bracket) instead of the standard one $\backslash b\backslash w\backslash w+\backslash b$ (matching any word only).

\subsubsection{Data \& Model Selection}
We decided to create eight small models, each corresponding to one activity. This made it possible to transform the multi-label classification problem into a binary classification problem, where each model only needs to check whether the cell belongs to the corresponding activity that the model classifies.

In addition, we use the \textit{XGBoost Classifier}, which uses gradient-boosting and is a widely used algorithm for classification tasks \cite{chen2016xgboost}. Further, a DT-based algorithm worked better than neural networks since we only used a small training sample. Through extensive grid-testing, we determined learning rates for each model separately.

As a basis for the training and evaluation datasets, we used notebooks from KGTorrent, a dataset of Python notebooks from Kaggle, crafted by Quaranta et al. \cite{Quaranta_2021}. It includes 248761 notebooks, spanning a period from November 2015 to October 2020. We extracted a set of 1000 notebooks from which we selected 2504 code cells. These cells act as our training \& validation dataset ($d^{train,val}$). Since the extracted notebooks contain an average of 23 cells, the 2504 code cells correspond to approximately 109 notebooks. Furthermore, we extracted 120 notebooks as an evaluation dataset ($d^{eval}$). We took care that the datasets $d^{train,val}$ and $d^{eval}$ were disjunct. This way, our evaluation results are not distorted since the DT models had never seen the code cells. Furthermore, we selected the notebooks randomly to replicate real-world performance most authentically.

To get correctly classified datasets, we labeled the selected notebooks with a preliminary version of \textsc{JupyLabel} and manually corrected all mistakes. The cells of each dataset were merged into one JSON file for the $d^{train,val}$ dataset and one JSON file for the $d^{eval}$ dataset. We denote the manually corrected datasets as \textit{source of truth}.

\subsubsection{Model Training and Validation Data}
\label{sec:training_data}

We split the $d^{train,val}$ dataset into separate training ($d^{train}$) and validation ($d^{val}$) datasets. The $d^{train}$ dataset consists of approximately 87 notebooks, while the $d^{val}$ dataset consists of 22 notebooks. 

All eight DT models were then trained on the code cells of the $d^{train}$ dataset, which equals around 80\%. The remaining 22 notebooks were used for validation after training. Splitting a dataset into an 80/20 training and validation set is a common practice in ML system development \cite{joseph2022optimal}.

Due to the inherent structure of notebooks, where numerous cells are present but only a small subset of them correspond to a specific activity, a significant \textit{class imbalance} arises, i.e., most often, a model predicts a negative outcome, given that there are many more cells unrelated to the specific activity it is trained to detect than those that are related. To account for these imbalances during training, we used \textit{resampling} techniques \cite{scikitResample}, which slightly helped to increase the performance metrics. However, resampling did not help to completely solve the problem of class imbalances, as already stated in Pecorelli et al. (2019) \cite{pecorelli2019role}.

\subsubsection{Model Evaluation Data}
Using the notebooks contained in the $d^{eval}$ dataset, we analyzed the distribution of labels per notebook (see Table \ref{tab:8.7}). Since this is a multi-label classification problem, cells can have none, one or more labels, resulting in this distribution. The numbers show similarities with the analysis by Ramsamy et al. (2023) \cite{ramasamy2023workflow}. As can be seen, the \textit{validate data} activity is the most prominent one, followed by the activity \textit{process data}. However, we encounter some differences when considering activities that are not dominant in a notebook. This can be justified by the difference in the dataset size and the selection of notebooks. We selected 120 notebooks, contrasting Ramsamy et al., who chose 500 notebooks through manual inspection. As a result, our $d^{eval}$ dataset comprises notebooks that may not adhere to best practices and might be incomplete concerning  ML activities. We intended to ensure that our cell classification approach was flexible enough to handle notebooks created by both experts and beginners.

\begin{table}[]
\caption{Label distribution in the evaluation dataset $d^{eval}$}
\label{tab:8.7}
\centering
\begin{tabular}{ll}
\hline
\textit{Activity}& \textit{\begin{tabular}[c]{@{}l@{}}Average distribution \\ per notebook in \%\end{tabular}}            \\ \hline
validate data      & 43.39   \\
process data       & 36.25   \\
setup notebook     & 21.49   \\
visualize data     & 16.68   \\
train model        & 15.53    \\
evaluate model     & 15.53    \\
ingest data        & 10.00   \\
transfer results   & 5.55    \\ \hline
Not labeled cells  & 1.00    \\ \hline
\end{tabular}
\end{table}

\section{JupyLabel }
\label{sec:JupyLabel}
The designed classifiers are implemented in a new cell classification tool called \textsc{JupyLabel}. In the following, we introduce its architecture and the workflow of its components to classify the cells. Then, we describe the central components in detail and finally give some information about the implementation of \textsc{JupyLabel}.

\subsection{Architecture}
\label{sec:arc}
\begin{figure*}
    \centering
    \includegraphics[width=\textwidth]{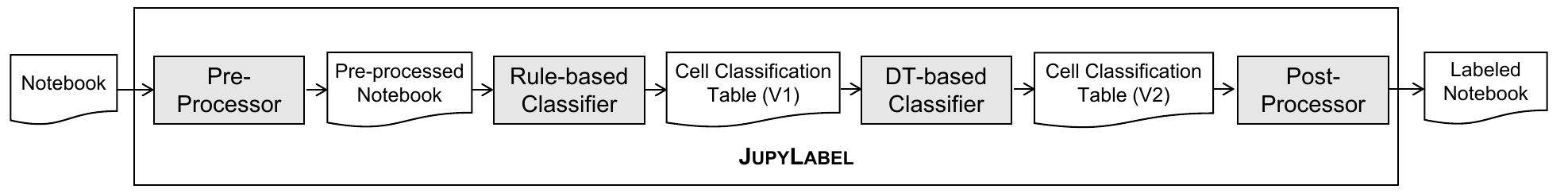}
    \caption{Architecture of \textsc{JupyLabel}}
    \label{fig:JupyLabel-architecture}
\end{figure*}

Figure \ref{fig:JupyLabel-architecture} shows the designed software architecture of \textsc{JupyLabel}. This architecture corresponds to the pipe-filter architecture pattern consisting of the four filter components (1) \textit{Pre-Processor,} (2) \textit{Rule-based Classifier}, (3) \textit{DT-based Classifier}, and (4) \textit{Post-Processor}.

As illustrated, a notebook is first prepared for further processing by the \textit{pre-processor}. Then, the \textit{pre-processed notebook} is given to the \textit{rule-based classifier} that applies a set of rules to classify the cells. This results in a \textit{cell classification table} containing metadata for each cell, e.g., output and cell type, and the determined labels. Cells that the rule-based classifier cannot classify are then passed to the \textit{DT-based classifier} that completes the cell classification table. Finally, the \textit{post-processor} inserts header annotations (the labels) into the notebook based on the cell classification table, resulting in a \textit{labeled notebook}.

\begin{figure}
    \centering
    \includegraphics[width=\linewidth]{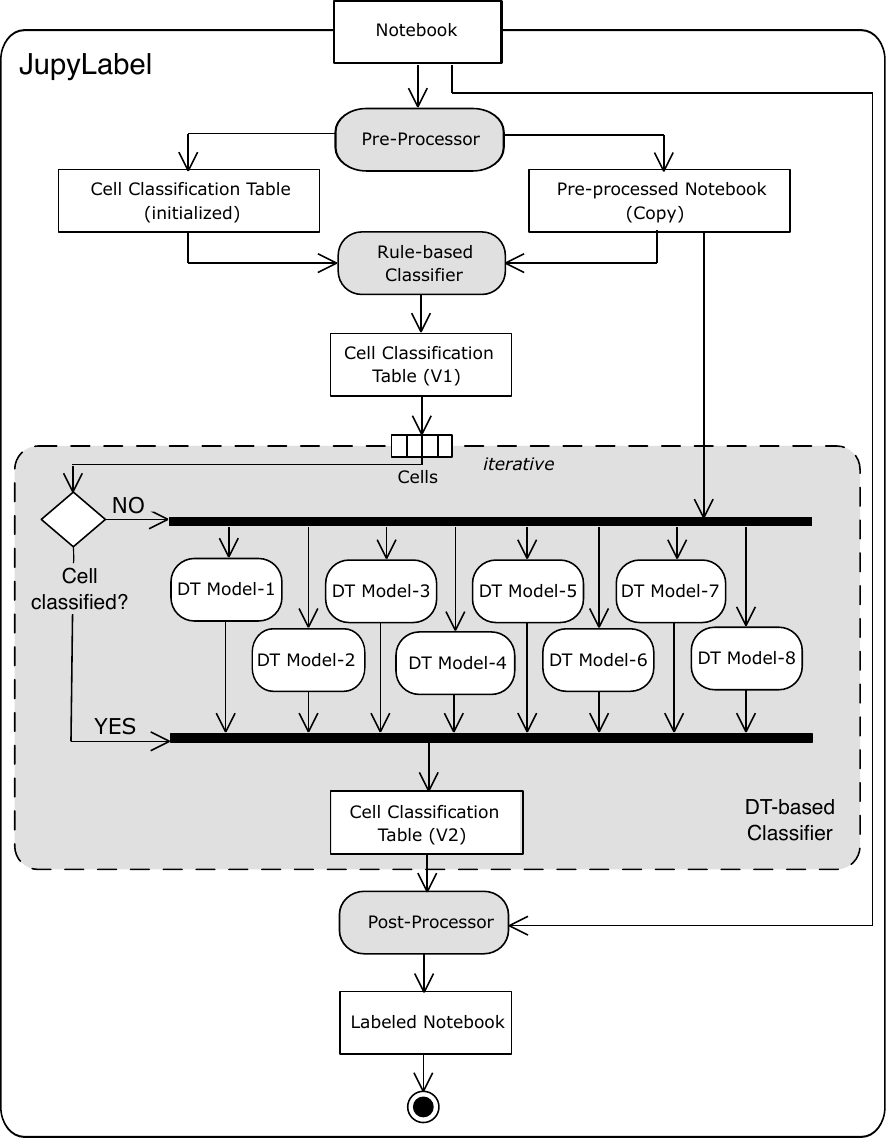}
    \caption{Detailed workflow (UML activity diagram)}
    \label{fig:JupyLabel-workflow}
\end{figure}
Figure \ref{fig:JupyLabel-workflow} shows the detailed workflow between the components. As can be seen, all cells that the rule-based classifier cannot classify are given to the DT-based classifier, which tries to classify them with the help of the eight DT models, one for each activity.

\subsection{The Pre-Processor}
The task of the pre-processor is to remove or rename cell contents that are not needed to classify a cell or could even lead to incorrect classification results.

The following pre-processing steps are performed to remove not needed contents:
\begin{itemize}
    \item\textit{Delete comments, newlines, and blank lines:} Often, comments describe what happens in subsequent cells and can confuse cell classification. Furthermore, all newlines and blank lines are deleted.
    \item\textit{Clear print statements:} Usually, print statements do not contain valuable hints to identify the activity performed in a cell. Therefore, all strings in print statements are deleted.
    \item \textit{Clear paths:} Paths are usually declared at the beginning of a notebook in a \textit{setup} activity cell. However, because they often contain words such as \smallttt{train}, \smallttt{validation}, or \smallttt{test}, they can give the impression that the code cell belongs to a model training or evaluation activity. To prevent this risk of confusion, all path identifiers are replaced by the keyword \smallttt{PATH}.
\end{itemize}
Additionally, these pre-processing steps are performed to facilitate the classification based on the heuristics presented in Section \ref{sec:design-RBC}:
\begin{itemize} 
    \item\textit{Rename setup code statements:} All import statements, comments, and magic commands are replaced by the keyword \smallttt{SETUP}. 
    \item\textit{Replace implicit returns:} Cells always return the value of the last evaluation expression if nothing was printed to the console. Therefore, all last lines in a code cell that access an attribute of an object such as \smallttt{df.columns} or simply an object \smallttt{df} itself are replaced by the keyword \smallttt{VALIDATION}.
\end{itemize}
As a side effect, pre-processing reduces the overall size of the data but preserves all information relevant to cell classification. Furthermore, pre-processing can significantly boost the performance in classification problems \cite{uysal2014impact}.

\subsection{The Rule-based Classifier}
The rule-based classifier applies the heuristics presented in Section \ref{sec:design-RBC}  to the given cells and inserts corresponding labeling information in the cell classification table. 
  
\subsection{The DT-based Classifier}
This classification component implements the eight DT models used to classify cells that cannot be classified based on the defined rules.

The DT models are given a pre-processed code cell if the previously mentioned rule-based classifier did not provide a positive result. Each DT model will then produce a value of either $0$ or $1$, where $0$ denotes negative and $1$ denotes positive. This information is then saved in the cell classification table.

\subsection{The Post-Processor}
This component labels the cells of the notebook based on the cell classification table. It provides the option to either insert tags or Markdown cells with the corresponding label.

\subsection{Implementation}
\textsc{JupyLabel}, implemented in Python, is available both as a CLI tool\footnote{url: \url{https://github.com/m1guelperez/jupylab_cli}} and a JupyterLab extension\footnote{url: \url{https://github.com/m1guelperez/jupylab_ext}}. The CLI offers multiple configurable flags. Notably, when using the CLI, users can choose between labeling the notebook with header annotations or using tags that insert labels directly into the cell headlines. Additionally, the CLI can be used in \textit{debug mode}, whereby it prints detailed information on the classification progress.

% It can be obtained from \textsc{Anonymous GitHub}\footnote{url: \url{https://pypi.org/}}. 

The developed DT models are available for offline or local use and are completely open-source. Thus, \textsc{JupyLabel} can be used for privacy-sensitive applications or environments. 

\begin{table*}[!htbp]
\caption{Metric scores of \textsc{JupyLabel} applied on the evaluation dataset $d^{eval}$}
\label{tab:metrics-120-nb}
\centering
\begin{tabular}{lllll}
\hline
\textit{Activity} & \textit{Accuracy} & \textit{Average Macro Precision} & \textit{Average Macro Recall} & \textit{Average Macro F1-Score} \\ \hline
setup notebook    & 0.9987            & 0.9982                           & 0.9971                        & 0.9976                          \\
ingest data       & 0.9891            & 0.9882                           & 0.9338                        & 0.9591                          \\
validate data     & 0.9392            & 0.9369                           & 0.9391                        & 0.9379                          \\
process data      & 0.9166            & 0.8992                           & 0.9174                        & 0.9071                          \\
train model       & 0.9761            & 0.9513                           & 0.9099                        & 0.9293                          \\
evaluate model    & 0.9657            & 0.9011                           & 0.8689                        & 0.8842                          \\
transfer results  & 0.9935            & 0.8968                           & 0.9410                        & 0.9177                          \\
visualize data    & 0.9887            & 0.9900                           & 0.9785                        & 0.9841                          \\ \hline
\textbf{Average}  & \textbf{0.9710}   & \textbf{0.9452}                  & \textbf{0.9357}               & \textbf{0.9396}                 \\ \hline
\end{tabular}
\end{table*}

\section{Evaluation}
\label{sec:eval}
This section provides information about the performance of the proposed flexible cell classification approach, particularly its correctness. We also provide measures to show how fast the classification works in real-world scenarios.

In the following, we used \textit{macro averaging} for all metrics to account for the data disparity regarding the class imbalance, as mentioned in \ref{sec:training_data}. This allows us to obtain a more balanced insight into performance across classes, irrespective of their prevalence, as it avoids overemphasizing the more prominent class. This enables us to provide worst-case estimates for our achieved scores.

\subsection{Classification Performance}
We used an unlabeled copy of our $d^{eval}$ dataset of 120 notebooks to evaluate the correctness and again applied \textsc{JupyLabel}. After that, we compared the labeled notebooks with our established source of truth and assessed the most common ML evaluation metrics like \textit{F1-score}, \textit{accuracy}, \textit{precision}, and \textit{recall}. 
We eliminated duplicated cells to ensure the repeated labeling of identical content did not skew our results. Furthermore, we excluded cells that contained code designed to interface specifically with the Kaggle environment, such as \smallttt{q1.hint()} or \smallttt{s2.step()}.

The metric scores obtained using \textsc{JupyLabel} on our $d^{eval}$ dataset are listed in Table \ref{tab:metrics-120-nb}. For the sake of completeness, we also calculated the score of each metric if we did not filter out duplicated cells. However, including duplicated cells only increased \textit{accuracy, precision, recall} and \textit{F1-score} by an average of 0.27\,\text{\%}, which is negligible in our case.

\subsection{Impact of Regular Expression}
We evaluated the impact of changing the regular expression of the \textit{CountVectorizer}.
As shown in Table \ref{tab:regex-impact}, it can be observed that especially \textit{precision, recall,} and \textit{F1-score} benefited from the improved regular expression.

\begin{table}[h!]
\caption{Impact of the improved regular expression}
\label{tab:regex-impact}
\begin{tabular}{llll}
\hline
\textit{Metric}         & \textit{Default regex} & \textit{Improved regex} & \textit{Change} \\ \hline
Accuracy                & 0.963                  & 0.971                   & +0.83\,\%  \\
Macro Average Precision & 0.933                  & 0.945                   & +1.29\,\%  \\
Macro Average Recall    & 0.916                  & 0.936                   & +2.18\,\%    \\
Macro Average F1-Score  & 0.923                  & 0.940                   & +1.84\,\%    \\ \hline
\end{tabular}
\end{table}

When investigating further the impact of the improved regular expression on the classification of positive samples, an even higher increase can be observed, as shown in Table \ref{tab:impact-pos-ex}. This is extremely important because of the huge class imbalances, as mentioned earlier. Increasing the performance of detecting actual positive samples by such a significant margin makes our approach much more applicable in real-world scenarios because it can detect activities and classify cells correctly even when facing class imbalances.

\begin{table}[h]
\caption{Impact of the improved regular expression to positive samples}
\label{tab:impact-pos-ex}
\begin{tabular}{llll}
\hline
\textit{Metric}                                                                        & \textit{Default regex} & \textit{Improved regex} & \textit{Change} \\ \hline
\begin{tabular}[c]{@{}l@{}}Macro Average Precision\\ for positive samples\end{tabular} & 0.898                  & 0.910                   & +1.36\,\%         \\
\begin{tabular}[c]{@{}l@{}}Macro Average Recall\\ for positive samples\end{tabular}    & 0.854                  & 0.893                 & +4.57\,\%         \\
\begin{tabular}[c]{@{}l@{}}Macro Average F1-Score\\ for positive samples\end{tabular}  & 0.872                  & 0.900                   & +3.21\,\%         \\ \hline
\end{tabular}
\end{table}

When comparing these results with the results obtained using a version of \textsc{JupyLabel} that did not use the rule-based classifier, the macro average metrics increased slightly (\textit{recall} increased by 0.76\,\text{\%},  \textit{precision} increased by 0.18\,\text{\%} and  \textit{F1-score} increased by 0.50\,\text{\%}), while the runtime performance remained unchanged.

\subsection{Runtime Performance}
We evaluated \textsc{JupyLabel} on an Apple M2 Max 12-Core computer with 32\,\text{GB} of RAM and running MacOS Sonoma 14.1. The dataset consists of 1000 notebooks with an average of 32 cells. We used the  \textsc{JupyLabel} CLI for this evaluation, labeled the dataset 10 times, and then calculated the average runtime.

The measured total execution time shows that \textsc{JupyLabel} takes 0.074\,\text{s} per notebook on average, highlighting its suitability in real-world scenarios, even when multiple intermediate files are required. Another reason for this performance is that our models and vectorizers are very compact and do not require powerful machines for inference. The vectorizers have an average size of 38\,\text{KB}, while the models have an average size of 156\,\text{KB}.

\subsection{Comparison with \textsc{HeaderGen}}
\label{sec:comparison}
In this section, we compare \textsc{JupyLabel} with \textsc{HeaderGen}, a current state-of-the-art header generation tool developed by Venkatesh et al. (2023) \cite{venkatesh2023enhancing}. For reproducibility reasons, we used the dataset consisting of 15 notebooks ($hg^{eval}$) provided by the authors. 

Before comparing both tools, we need to point out the differences in labeling. \textsc{HeaderGen} classifies cells according to the following five activity labels: \textit{Library Loading, Visualization, Data Processing and Exploration, Feature Engineering,} and \textit{Model Building and Training}. Because our approach defines a finer and more extensive set of activity labels, we faced the challenge of creating the fairest possible comparison that accounts for these differences.

To accomplish this, we manually labeled all 15 notebooks of the $hg^{eval}$ dataset using our set of labels to create a source of truth. We then compared their source of truth, which considers their five labels, with our newly created source.  Since we did not find any significant semantic differences, apart from the fact that our labels are more fine-grained, we can compare the two approaches apart from these differences.

Table \ref{tab:tool-HG-NBs-compact} shows the metric scores obtained by \textsc{JupyLabel} when classifying these 15 notebooks.

\begin{table*}[]
\centering
\caption{Metric scores using \textsc{JupyLabel} on the $hg^{eval}$ dataset}
\label{tab:tool-HG-NBs-compact}
\begin{tabular}{lllll}
\hline
\textit{Activity} & \textit{Accuracy} & \textit{\begin{tabular}[c]{@{}l@{}}Average Macro \\ Precision\end{tabular}} & \textit{\begin{tabular}[c]{@{}l@{}}Average Macro \\ Recall\end{tabular}} & \textit{\begin{tabular}[c]{@{}l@{}}Average Macro \\ F1-Score\end{tabular}} \\ \hline
setup notebook    & 1.0               & 1.0                                                                         & 1.0                                                                      & 1.0                                                                        \\
ingest data       & 1.0               & 1.0                                                                         & 1.0                                                                      & 1.0                                                                        \\
validate data     & 0.9656            & 0.9638                                                                      & 0.9649                                                                   & 0.9643                                                                     \\
process data      & 0.9625            & 0.9543                                                                      & 0.9670                                                                   & 0.9599                                                                     \\
train model       & 0.9906            & 0.9818                                                                      & 0.9697                                                                   & 0.9756                                                                     \\
evaluate model    & 0.9844            & 0.9740                                                                      & 0.9293                                                                   & 0.9503                                                                     \\
transfer results  & 0.9906            & 0.9728                                                                      & 0.9529                                                                   & 0.9626                                                                     \\
visualize data    & 0.9906            & 0.9950                                                                      & 0.9250                                                                   & 0.9570                                                                     \\ \hline
\textbf{Average}  & \textbf{0.9855}   & \textbf{0.9802}                                                             & \textbf{0.9634}                                                          & \textbf{0.9712}                                                            \\ \hline
\end{tabular}
\end{table*}

According to \cite{venkatesh2023enhancing}, \textsc{HeaderGen} achieved a recall of 96.8\,\text{\%}, a precision of 82.2\,\text{\%}, and a F1-score of 88.9\,\text{\%} in classifying these notebooks. We were able to locally reproduce these results, using the most recent \textsc{GitHub} version of \textsc{HeaderGen}\footnote{URL: https://github.com/sergeychernyshev/HeaderGen, revision: aa9f883}. When comparing these scores with the corresponding ones of \textsc{JupyLabel}, an improvement of 19.25\,\text{\%} regarding precision, and 9.25\,\text{\%} for the F1-score can be observed. However, the recall score dropped by 0.45\,\text{\%}.

It is important to note that the 15 notebooks of the $hg^{eval}$ dataset were not executed; therefore, their cells had no output type. Considering this, the metric scores for the activity \textit{visualize data} would probably be slightly higher or even reach the perfect score, as in the case of the activity \textit{setup notebook}, due to the rule-based classifier used.
 
Furthermore, we executed the most recent \textsc{HeaderGen} version on the same Apple M2 Max computer on which we executed \textsc{JupyLab}. \textsc{HeaderGen} took 10.99\,\text{s} to label the cells of the 15 notebooks, while \textsc{JupyLabel} took only 1.42\,\text{s}, making it 87.08\,\text{\%} faster. To compare the execution time of \textsc{JupyLab} and \textsc{HeaderGen} as fair as possible, we executed both in a Docker container with the same environment.

\subsection{Summary}
With an average accuracy of 97.10\,\text{\%}, a macro average precision of 94.52\,\text{\%}, a macro average recall of 93.57\,\text{\%}, and a macro average F1-score of 93.96\,\text{\%}, \textsc{JupyLabel} achieved excellent scores on the $d^{eval}$ dataset. Since our evaluation covers a wide range of notebooks, including notebooks created by beginners that are far more difficult to label correctly, it is also suitable for complex notebooks.

\textsc{JupyLabel} does its job very quickly, taking on average only 0.074\,\text{s} to label one notebook. This makes it suitable for real-world applications, as our performance scores clearly show.

\begin{figure}
    \centering
    \includegraphics[width=\linewidth]{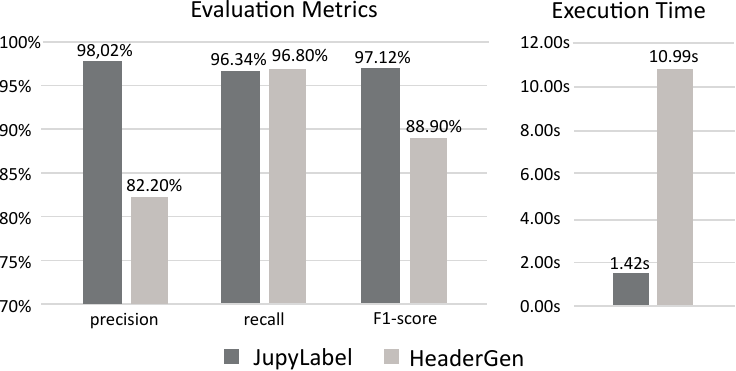}
    \caption{Comparison of the metric scores}
    \label{fig:tool-headergen-comp}
\end{figure}

Additionally, \textsc{JupyLabel} outperforms the state-of-the-art tool \textsc{HeaderGen} regarding precision, and F1-score as well as execution time (see Figure \ref{fig:tool-headergen-comp}). 

\section{Conclusion \& Future Work}
\label{sec:conclusion}
The main contribution presented in this paper is a new approach to cell classification of ML projects in notebooks. This approach is highly flexible because its classifiers can easily adapt to a changing environment. This is achieved through a hybrid classification approach combining a rule-based and a DT-based classifier, whereby the latter can be retrained. 

The evaluation results show that the developed tool \textsc{JupyLabel}, which implements this classification approach, provides excellent metric results regarding precision, recall, and F1-score. In addition, we could show that \textsc{JupyLabel} outperforms the current state-of-the-art tool \textsc{HeaderGen}. In the future, we plan to do research in the following areas:

With its high metric scores, it is very suited to support navigation in notebooks better. Exploring the collected tracking information would be much easier if our cell classification approach were integrated into JupyterLab extensions for notebook tracking, e.g., \textsc{Verdant} developed by Kery et al. \cite{mkery2018}.

Moreover, our cell classification approach can be used as a clustering method to group cells by ML activities. This would allow us to mine information on how data scientists work from a large set of notebooks.

It was challenging to evaluate and compare existing approaches for the objective classification of cells. Therefore, the dataset we used for evaluation could be extended with further elements defined by Tichy \cite{tichy2014}, resulting in a publicly usable benchmark for cell classification.

Finally, we will do more research in using generative AI systems, like \textsc{ChatGPT}, to the given cell classification approach. We conducted first experiments, which led to promising results but also revealed new challenges.

% Put references at the bottom
\bibliographystyle{IEEEtran}
\bibliography{Bibliography}
\end{document}